\documentclass[showpacs,preprintnumbers,amsmath, prd]{revtex4}
\begin{document}

\title{Quasinormal modes of Unruh's Acoustic Black Hole}
\author{Joel Saavedra}
\email{Joel.Saavedra@ucv.cl}
\affiliation{Instituto de F\'{\i}sica, Pontificia Universidad Cat\'olica de Valpara\'{\i}%
so, \\
Casilla 4950, Valpara\'{\i}so, Chile.}

\begin{abstract}
We have studied the sound perturbation of Unruh's acoustic
geometry and we present an exact expression for the quasinormal
modes of this geometry. We are obtain that the quasinormal
frequencies are pure-imaginary, that give a purely damped modes.

\end{abstract}

\maketitle
\section{\label{sec:Int}Introduction}
General relativity has many successful predictions, some of them
are passing experimental tests with high precision, for example
mercury's perihelion  and gravitational redshift. On the other
hand, the most spectacular and promising prediction is the
existence of black holes. In this sense the theoretical area of
black hole physics became an attractive and productive field of
research in the past century. Those classical and quantum
properties are well understood within the general relativity
framework. From a classical point of view, nothing can escape from
the black holes; however, from the quantum point of view, black
holes are not completely black and they can emit radiation with a
temperature given by $h/8 \pi k_{B}GM$ \cite{Hawking:1974sw}. This
radiation is essentially thermal and they slowly evaporate by
emitting quanta. Another property of interest that play an
important role in black hole physics are the quasinormal modes
(QNM's). They determine the late-time evolution of fields in the
exterior of the black hole, and numerical simulations of stellar
collapse and black hole collisions have shown that in the final
stage of such processes the quasinormal modes eventually dominate
the black hole response to any kind of perturbation. QNM's has
received great attention in the last few years, since it is
believed that these models shed light on the solution or
understanding  to fundamental problems in loop quantum gravity. In
particular, from Hod's proposal \cite{Hod:1998vk}, modes with high
damping have received great attention, specially their relevance
in the quantization of the area of black holes
\cite{Setare:2004uu,Choudhury:2003wd, Padmanabhan:2003fx} and the
possibilities of fixing the Immirzi parameter in loop quantum
gravity \cite{Dreyer:2002vy, Kunstatter:2002pj, Natario:2004jd}.

Obviously the experimental test are impossible to make at the
level of terrestrial laboratories. This opens the interest in
analog models that mimic the properties of black hole physics. In
this sense, the field of analog models of gravity allows in
principle that the most processes of black hole physics can be
studied in a laboratory. In particular the use of supersonic
acoustics flows as an analogy to gravitating systems was for first
time proposed by Unruh \cite{Unruh:81} and with the works of
Visser \cite{Barcelo:2005fc, Novello:2002qg, Visser:1998qn,
Visser:1997ux, Visser:1993ub} has received an exponentially
growing attention.

The basis of the analogy between gravitational black holes and
sonic black holes comes from considering the propagation of
acoustic disturbances on a barotropic, inviscid, inhomogeneous and
irrotational (at least locally) fluid flow.  It is well known that
the equation of motion for this acoustic disturbance (described by
its velocity potential ) is identical to the Klein-Gordon equation
for a massless scalar field minimally coupled to gravity in a
curved spacetime \cite{Visser:1998qn, Visser:1997ux,
Visser:1993ub}. QNM's modes for acoustic black holes in 2+1
dimension was computed in Refs. \cite{Cardoso:2005ij,
Cardoso:2004fi, Lepe:2004kv,Berti:2004ju} and for analogous black
holes in Bose-Einstein condensate in Ref. \cite{Nakano:2004ha}. In
this work we analytically compute the QNMs or acoustic
disturbances in Unruh's 3+1 dimensional sonic black hole (Laval
nozzle fluid flow), and also we show their large damping limit.

The organization of the paper is as follows: In Sec. II we specify
the Unruh's Sonic Black Hole. In Sec. III we determine the QNMs,
and the large damping limits. Finally, we conclude in Sec. V.

\section{\label{sec:Sonicbh}Unruh's Sonic Black Hole}

The idea of using supersonic acoustic flows as analog systems to
mimic some properties of black hole physics was proposed for the
first time by Unruh \cite{Unruh:81}. Essentially he showed the
possibility to use sonic flow in order to explore properties like
the Hawking temperature near the sonic horizon and, in principle
to developed an experimental setting to study the fundamental
problem of evaporation of real general relativity black holes.As
we mentioned in the introduction, the basis of the analogy between
gravitational black hole and sonic black holes comes from
considering the propagation of acoustic disturbances on a
barotropic, inviscid, inhomogeneous and irrotational (at least
locally) fluid flow. \ It is well known that the equation of
motion for this acoustic disturbance (described by its velocity
potential $\psi $) is identical to the Klein-Gordon equation for a
massless scalar field minimally coupled to gravity in a curved
space \cite{Visser:1998qn, Visser:1997ux, Visser:1993ub},

In Unruh's work, the acoustic geometry is described by the
following sonic line element

\begin{equation}
ds^{2}=\frac{\rho _{0}}{\tilde{c}}\left( -\left( \tilde{c}%
^{2}-v_{o}^{r2}\right) d\tau ^{2}+\frac{\tilde{c}dr^{2}}{\tilde{c}%
^{2}-v_{o}^{r2}}+r^{2}d\Omega ^{2}\right) ,  \label{eq1}
\end{equation}%
where $\rho _{0\text{ }}$ is the density of the fluid, $\tilde{c}$
is the velocity of sound in the fluid (by simplicity we will
assume these quantities constant) and $v_{0}^{r}$ represent the
radial component of the flow velocity.

On the other hand, if then we assume that at some value of
$r=r_{+}$ we have the background fluid smoothly exceeding the
velocity of sound,

\begin{equation}
v_{0}^{r}=-\tilde{c}+\tilde{a}(r-r_{+})+\vartheta (r-r_{+})^{2},  \label{eq2}
\end{equation}%
the above metric assumes just the form it has for a Schwarzschild
metric near the horizon. In \ this limit metric our (\ref{eq1})
reads as follows

\begin{equation}
ds^{2}=\frac{\rho _{0}}{\tilde{c}}\left( -2\tilde{a}\tilde{c}(r-r_{+})d\tau ^{2}+%
\frac{\tilde{c}dr^{2}}{2\tilde{a}\tilde{c}(r-r_{+})}+r^{2}d\Omega
^{2}\right) , \label{eq3}
\end{equation}%
where $\tilde{a}$ is a parameter associated with the velocity of
the fluid defined as $(\nabla\cdot\overrightarrow{v})|_{r=r_{+}}$
\cite{Kim:2004sf}. Note that this geometry was study in Ref.
\cite{Kim:2004sf} where the author studied the low energy dynamics
and obtained the greybody factors for the sonic horizon from the
absorption and the reflection coefficients.

In the quantum version of this system we can hope that the
acoustic black hole emits ''acoustic Hawking radiation''. This
effect coming from the horizon of events is a pure kinematical
effect that occurs in any Lorenzian geometry independent of its
dynamical content \cite{Visser:1997ux}. It is well known that the
acoustic metric does not satisfied the Einstein equations, due to
the fact that the background fluid motion is governed by the
continuity and the Euler equations. As a consequence of this fact,
one should expect that the thermodynamic description of the
acoustic black hole is ill defined. However, this powerful analogy
between black hole physics and acoustic geometry admit to extend
the study of many physical quantities associated to black holes,
such as quasinormal modes which we consider in the next section.

\section{\label{sec:Sonicbh1}Quasinormal modes of Unruh's Sonic Black Hole}

The basis of the analogy between Einstein black holes and sonic
black holes comes from considering the propagation of acoustic
disturbances on a barotropic, inviscid, inhomogeneous and
irrotational fluid flow. It is well known that the equation of
motion for these acoustic disturbances (described by its velocity
potential $\Phi $) is identical to the Klein-Gordon (KG) equation
for a massless scalar field minimally coupled to gravity in a
curved spacetime \cite{Visser:1998qn, Visser:1997ux,
Visser:1993ub},

\begin{equation}
\frac{1}{\sqrt{g}}\partial _{\mu }\left( \sqrt{g}g^{\mu \nu }\partial _{\nu
}\Phi \right) =0.  \label{kg1}
\end{equation}
In order to compute the QNM's we apply the standard procedure
described in
Refs.\cite{Kim:2004sf}\cite{Birmingham:2001hc}\cite{Fernando:2003ai},
and begin to rewrite the metric (\ref{eq1}) in the following form

\begin{equation}
ds^{2}=-f(r)d\tau ^{2}+\frac{\tilde{c}dr^{2}}{f(r)}+r^{2}d\Omega ^{2},
\label{eq4}
\end{equation}
where $f(r)=2\tilde{a}\tilde{c}(r-r_{+})$ \ In order to compute
the QNM'S we need to solve the Klein Gordon equation (\ref{kg1})
in curve space described by the metric (\ref{eq4}) and, by virtue
of symmetries of the metric we use the following Ansatz for the
scalar field

\[
\Phi =e^{-i\omega t}Y_{l}^{m}(\theta ,\varphi )R(r),
\]%
with this ansatz the KG equation can be separated as follows

\begin{eqnarray}
\frac{1}{\sin \theta }\left( \partial _{\theta }\left( \sin \theta \partial
_{\theta }Y_{l}^{m}(\theta ,\varphi )\right) +\frac{1}{\sin \theta }\partial
_{\varphi }^{2}Y_{l}^{m}(\theta ,\varphi \right) &=&l(l+1)Y_{l}^{m}(\theta
,\varphi ),  \label{eq5} \\
\frac{1}{\tilde{c}^{2}r^{2}}\frac{d}{dr}\left( r^{2}f(r)\frac{d}{dr}%
R(r)\right) +\left( \frac{\omega ^{2}}{f(r)}+\frac{l(l+1)}{r^{2}}\right)
R(r) &=&0.  \nonumber
\end{eqnarray}%
Now, if then we consider the change of variables $z=1-r_{+}/r$,
that transform the radial equation to

\begin{equation}
z(1-z)\frac{d^{2}R}{dz^{2}}+\frac{dR}{dz}+P(z)R=0,  \label{eq6}
\end{equation}%
here,

\begin{equation}
P(z)=\frac{B}{z(1-z)}-A,  \label{p}
\end{equation}%
where,

\begin{eqnarray}
A &=&\frac{l(l+1)\tilde{c}}{2\tilde{a}r_{+}},  \label{eq7} \\
B &=&\left( \frac{\omega }{2\tilde{a}}\right) ^{2}.  \nonumber
\end{eqnarray}

Note that in the new coordinate system, $z=0$ correspond to the
horizon and $z=1$ corresponds to infinity, and if we consider the
definition

\begin{equation}
R=z^{\alpha }(1-z)^{\beta }F(z),  \label{eq8}
\end{equation}%
then the radial equation satisfies the standard hypergeometric
form \cite{abramowitz}

\begin{equation}
z(1-z)\frac{d^{2}F}{dz^{2}}+(c-(1+a+b)z)\frac{dF}{dz}+abF=0,  \label{eq9}
\end{equation}%
where

\begin{eqnarray*}
c &=&2\alpha +1, \\
ab &=&A+(\alpha +\beta )(\alpha +\beta -1), \\
a+b &=&2(\alpha +\beta )-1
\end{eqnarray*}%
and

\begin{eqnarray}
\alpha ^{2} &=&-B,  \label{eq10} \\
\beta &=&1\pm \sqrt{1-B}.  \nonumber
\end{eqnarray}%
Without loss of generality, we put $\alpha =-i\sqrt{B}$ and $\beta =1-\sqrt{%
1-B}.$ It is well known that the hypergeometric equation has three
regular singular point at $z=0,$ $z=1$ and $z=\infty $, and it has
two independent solutions in the neighborhoods of each point
\cite{abramowitz}. The solutions of the radial equation reads as
follows

\begin{equation}
R(z)=C_{1}z^{\alpha }(1-z)^{\beta }F(a,b,c,z)+C_{2}z^{-\alpha }(1-z)^{\beta
}F(a,b,c,z).  \label{eq11}
\end{equation}

In the gravitational case quasinormal modes of a scalar classical
perturbation of black holes are defined as the solution the
Klein-Gordon equation characterized by purely ingoing waves at the
horizon $\Phi \sim e^{-i\omega (t+r)}$, since at least classically
outgoing flux is not allowed at the horizon. In addition, one has
to impose boundary conditions on the solutions at the asymptotic
region (infinity). It is crucial using the asymptotic geometry of
the space time under study. In the case of asymptotically flat
space time, the condition we need to imposes to the wave functions
is a purely outgoing waves $\Phi \sim e^{-i\omega (t-r)}$ at the
infinity \cite{Horowitz:1999jd}. For non asymptotically flat space
time (AdS space time for example), several boundary conditions
have been discussed in the literature. In the two dimensional
cases of BTZ\ black holes, the quasinormal modes for scalar
perturbations were found in Refs.
\cite{Birmingham:2001hc}\cite{Cardoso:2001hn} by imposing the
vanishing Dirichlet condition at infinity: In Ref.
\cite{Birmingham:2001pj} these modes were computed with the
condition of a vanishing energy momentum flux density at
asymptotia. In this paper we consider the vanishing flux
condition.

In the neighborhood of the horizon $(z=0)$ using the property
$F(a,b,d,0)=1$ the radial solution is given \ by

\begin{equation}
R(z)=C_{1}\exp (-i\frac{\omega }{2\tilde{a}}\ln (1-\frac{r_{+}}{r})+C_{2}\exp (i%
\frac{\omega }{2\tilde{a}}\ln (1-\frac{r_{+}}{r})),  \label{eq12}
\end{equation}%
while the boundary conditions of purely ingoing waves at the horizon demands $%
C_{2}=0.$ In order to implement boundary condition at infinity
$(z=1)$, we use the linear transformation $z\rightarrow 1-z$
formula for the hypergeometric function and we obtain,

\begin{eqnarray}
R &=&C_{1}z^{\alpha }(1-z)^{\beta }\frac{\Gamma (c)\Gamma (c-a-b)}{\Gamma
(c-a)\Gamma (c-b)}F(a,b,a+b-c+1,1-z)+  \label{eq13} \\
&&+C_{1}z^{\alpha }(1-z)^{c-a-b+\beta }\frac{\Gamma (c)\Gamma (a+b-c)}{%
\Gamma (a)\Gamma (b)}F(c-a,c-b,c-a-b+1,1-z).  \nonumber
\end{eqnarray}

Using the condition that the flux is given by

\begin{equation}
\mathcal{F}=\frac{\sqrt{g}}{2i}(R^{\ast }\partial _{\mu }R-R\partial _{\mu
}R^{\ast }),  \label{eq14}
\end{equation}%
is vanished at infinity. Due to, $B>0$, the asymptotic flux has a
set of divergent \ terms, with the leading term of order
$(1-z)^{1-2\beta }$. Then according to Eq. (\ref{eq13}) each of
these terms are proportional to

\begin{equation}
\left\vert \frac{\Gamma (c)\Gamma (a+b-c)}{\Gamma (a)\Gamma (b)}\right\vert
^{2},  \label{eq15}
\end{equation}%
and hence for a vanishing flux at $z=1$ the following restrictions
have to be applied

\begin{equation}
a=-n\text{ or }b=-n,  \label{eq17}
\end{equation}

where $n=0,1,2,...$. These conditions lead directly to an exact
determination of the quasinormal modes as follows

\begin{equation}
\omega =-\frac{i}{2}\frac{(n-1)(n+3)\tilde{a}}{n+1}.  \label{eq18}
\end{equation}

Note that the quasinormal modes have the properties being purely
imaginary and independent of the radius of the horizon of the
sonic black holes. Besides this, it shows the instabilities of
this kind of analog black hole under sonic perturbations. Due to,
that pure imaginary frequency of the zero mode has the wrong sign
(positive)that mean an exponentially growing mode.

For the infinite limit damping (\textit{i.e.} the $n\rightarrow
\infty $ limit)

\[
\omega _{\infty }=-i(n+\frac{1}{2})\tilde{a}
\]

\section{\label{sec:Sonicbh4}Conclusions and Remarks}

In this paper we have computed the exact values of the quasinormal
modes of Unruh's sonic black holes and according to Ref.
\cite{Unruh:81} this QNM's should be  behaved similarly to QNM's
of near horizon Schwarzschild black holes. The QNM's are pure
imaginary, this kind of QNM's was reported in Refs.
 \cite{Lopez-Ortega:2005ep}\cite{Berti:2003ud}\cite{Fernando:2003ai}.
 Therefore, for black hole that show this kind of QNM's is impossible to compute the area spectrum
from the Kunstatter adiabatic invariant. On the other hand, the
QNM's formula show that the zero mode exhibit an exponentially
growing and therefore the Unruh's black hole in the limit that the
background fluid smoothly exceeding the velocity of sound and, for
the first order in the expansion (\ref{eq2}),  becomes one
instable geometry under sonic perturbations that excite the zero
In opposite direction this result shows the large stabilities of
the dumb hole for perturbations that excite the overtone with
$n>1$ and for all overtones in higher damping limit. We also want
to note that the frequencies of the QNM's do not depend of radius
of the horizon and on the angular momentum of the perturbation.
They only have a proportional dependence the control parameter
$(\tilde{a})$ which describes the velocity of the sonic flow, in
according to Ref. \cite{Kim:2004sf} where the decay rate for this
geometry and the thermal emission was only proportional to the
control parameter.

\begin{acknowledgments}
The author is grateful to C. Campuzano, S. Lepe, A. Lopez-Ortega,
R. Troncoso and E. Vagenas for many useful and enlightening
discussions. This work was supported by COMISION NACIONAL DE
CIENCIAS Y TECNOLOGIA through FONDECYT \ Postdoctoral Grant
3030025 . Was also were partially supported by PUCV Grant No.
123.778/05. The author wishes to thank the Centro de Estudios
Cient\'{\i}ficos (CECS) for its kind hospitality and U. Raff for a
careful reading of the manuscript.
\end{acknowledgments}

\end{document}